\documentclass[]{raa}            
\usepackage{graphicx,times}
\usepackage{natbib}

\providecommand{\bm}[1]{\mbox{\boldmath$#1$\unboldmath}}

\begin{document}

   \title{A Multicolour Photometric Study of the neglected eclipsing binary FT Ursae Majoris
}

 \volnopage{ {\bf 2009} Vol.\ {\bf 9} No. {\bf XX}, 000--000}
   \setcounter{page}{1}

   \author{Jin-Zhao Yuan}
   \institute{Department of Physics, Shanxi Normal University, Linfen
041004, Shanxi, China; {\it yuanjz@sxnu.edu.cn}\\}

\abstract{The multicolour photometric observations of the
neglected eclipsing binary FT Ursae Majoris (FT UMa) were obtained
in 2010. The 2003 version of Wilson-Devinney code was used to
analyze the light curves in $B$, $V$, and $R$ bands
simultaneously. Based on the spectroscopic mass ratio $q=0.984$
published by Pribulla et al., it is found that FT UMa is an
evolved contact binary with a contact degree of $15.3\%$. The low
amplitude of light variations, $\sim 0.15$ mag, arises mainly from
a moderately low inclination angle of $i=62.^{\circ}80$ and almost
identical components in size rather than the light dilution of a
third component, which contributes light of only $\sim 10\%$.
 }

   \authorrunning{Jinzhao Yuan}            
   \titlerunning{Multicolour Photometric study of FT UMa }  
   \maketitle


%
%
\section{Introduction}           
\label{sect:intro}

FT UMa is a relatively bright target with the maximum magnitude of
$V_{max}=9.25$. But, the variable star was not known until the
systematic Hipparcos survey took place because of its relatively
low amplitude of photometric variability, i.e., $\sim 0.15$ mag in
the $V$ band. First photometric study was carried by Ozavci et al.
(2007), who obtained the photometric mass ratio of
$q=0.25\pm0.01$. Pribulla et al. (2009), however, derived the mass
ratio, $q=0.984\pm0.019$, from the radial velocities of FT UMa.
So, the photometric study of the eclipsing binary FT UMa should be
re-carried out based on the spectroscopic mass ratio.

In this paper, the absolute physical parameters were determined
based on the CCD multicolour photometric observations. Two
interesting properties are discussed in the last section.


\section{Observations}
\label{sect:Obs}

New multicolour CCD photometric observations of FT UMa were
carried out on 2010 January 17, 20, and 21, and December 4 and 5
using the 85-cm telescope at the Xinglong Station of National
Astronomical Observatory of China (NAOC), equipped with a
primary-focus CCD photometer. The telescope provides a field of
view of about $16.^{'}5\times16.^{'}5$ at a scale of $0.^{''}96$
per pixel and a limit magnitude of about 17 mag in the $V$ band
(Zhou et al. 2009). The standard Johnson-Cousin-Bessel $BVR$
filters were used simultaneously. HD 233579 was selected as a
comparison star and J08532982+5123196 as a check star. The
coordinates are listed in Table 1.

The data reduction was performed by using the aperture photometry
package of IRAF{\footnote[1]{IRAF is developed by the National
Optical Astronomy Observatories, which are operated by the
Association of Universities for Research in Astronomy, Inc., under
contract to the National Science Foundation.}} (bias subtraction,
flat-field division). Extinction corrections were ignored as the
comparison star is very close to the variable. In total, 1597,
1579, and 1527 CCD images in the $B$, $V$, and $R$ bands were
obtained, respectively. Several new times of primary minimum are
derived from the new observation by using a parabolic fitting
method. Two times of light minima (i.e., HJD
$2455217.3290\pm0.0002$ and HJD $2455217.9832\pm0.0005$) were
obtained by averaging those in three bands.

The light curves are displayed in Figure 1. The orbital period
adopted to calculate the phase was taken from Pribulla et al.
(2009), i.e., 0.6547038 days. As shown in the bottom panel of
Figure 1, the differential magnitudes between the comparison star
and the check star varied within $\sim 0.04$ mag, which is due to
the inappropriate check star.

\begin{table}[h!!!]
\small \centering
\begin{minipage}[]{120mm}
\caption[]{ Coordinates of FT UMa and its Comparison and Check
Stars.}\label{Table 1}
\end{minipage}
\tabcolsep 6mm
\begin{tabular}{lll}
\hline\noalign{\smallskip}
Stars           & $\alpha_{2000}$         &$\delta_{2000}$ \\
\hline\noalign{\smallskip}
FT UMa          & $08^{h}54^{m}30.4^{s}$  & $51^\circ14'40.3''$\\
Comparison  & $08^{h}53^{m}31.5^{s}$ & $51^\circ26'47.2''$\\
Check       & $08^{h}53^{m}29.8^{s}$ & $51^\circ23'19.7''$\\
\noalign{\smallskip}\hline
\end{tabular}
\end{table}

\begin{figure}[h!!!]
\centering
\includegraphics[width=9.0cm,angle=0]{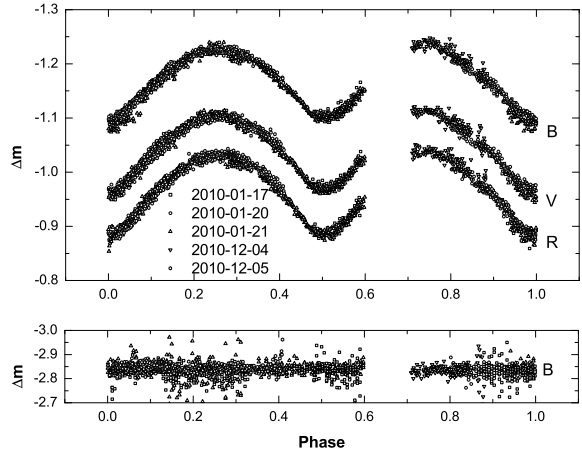}
\begin{minipage}[]{85mm}
\caption{Top panel: the light curves of FT UMa in the $B$, $V$,
and $R$ bands obtained on 2010 January 17, 20 and 21, and December
4 and 5. Bottom panel: the differential light curves of the
comparison star relative to the check star in the $B$ band.}
\end{minipage} \label{Fig1}
\end{figure}

\section{Photometric solutions with the W-D method}
As shown in Figure 1, the data show a nonuniform phase coverage.
The data averaged in phase with 0.01 phase bins were used
hereafter. The light curves were analyzed using the 2003 version
of the Wilson-Devinney code (Wilson \& Devinney 1971; Wilson 1979,
1990). In the process of solution, the spectroscopic mass ratio of
$q=0.984$ published by Rucinski et al. (2009) was fixed. Moreover,
Pribulla et al. (2009) gave a spectral type of F0. So an effective
temperature of $T_1 = 7178$K was assumed for the primary component
according to the table of Gray (2005). This is close to the
critical temperature of stars with convective and radiative
envelopes, i.e., 7200K. For security, both convective and
radiative envelopes were taken into account. The gravity-darkening
coefficients, $g_1=g_2=0.320$, were used for convective case and
$g_1=g_2=1.000$ for radiative case. The bolometric albedos,
$A_1=A_2=0.5$, were used for convective case and $A_1=A_2=1.0$ for
radiative case. The logarithmic limb-darkening coefficients came
from van Hamme (1993). The photometric parameters are listed in
Table 2.

As suggested by Pribulla et al. (2009), the photometric solution
started with mode 2 (i.e., detached mode). After some differential
corrections, the solution converged to mode 3 (i.e., overcontact
mode) in convective case, and to mode 5 (i.e., semi-detached mode)
in radiative case. All of the parameters derived from the model
are listed in Table 2. The the sum of the residuals squared for
convective case is smaller than that for radiative case. So, the
convective case is more plausible. The theoretical light curves
are plotted in Figure 2 as solid lines, which fit the observations
very well except for the deviations around the second maxima,
which are due to the relatively low quality of the points around
the second maxima.

Using the value of
$(M_{1}+M_{2})~\mathrm{sin}^3{}i=2.077~M_{\odot}$ (Pribulla et al.
2009), the following physical parameters can be derived:
$M_1=1.49(5)~M_{\odot}$, $R_1=1.79(2)~R_{\odot}$,
$L_1=7.68(19)~L_{\odot}$, $M_2=1.46(5)~M_{\odot}$,
$R_2=1.78(2)~R_{\odot}$, $L_2=6.86(22)~L_{\odot}$, and
$a=4.55(5)~R_{\odot}$. Given the mass-radius relation of
$R/R_{\odot}=(M/M_{\odot})^{0.73}$ and the mass-luminosity
relation of $L/L_{\odot}=1.2(M/M_{\odot})^{4.0}$ for ZAMS stars, a
main sequence star with a mass of $\sim1.49~M_{\odot}$ has a
radius of $\sim1.34~R_{\odot}$ and a luminosity of
$\sim5.91~L_{\odot}$, both of which are lower than those of two
components. We can conclude that FT UMa has evolved off the main
sequence.

\begin{table}
\caption{Photometric Solutions for FT UMa.}\label{tab1}
\begin{center}
\small
\begin{tabular}{lll}
\hline\noalign{\smallskip}
Parameters       &      Convective case     &     Radiative case          \\
\hline\noalign{\smallskip}
configure                   &overcontact                      & semi-detached \\
$g_{1}=g_{2}$               &0.32                             & 1.00      \\
$A_{1}=A_{2}$               &0.5                              & 1.0       \\
$x_{1bol}$                  &0.642                            & 0.642     \\
$x_{2bol}$                  &0.641                            & 0.641     \\
$y_{1bol}$                  &0.257                            & 0.257     \\
$y_{2bol}$                  &0.253                            & 0.243     \\
$x_{1B}  $                  &0.781                            & 0.781     \\
$x_{1V}  $                  &0.683                            & 0.683     \\
$x_{1R}  $                  &0.584                            & 0.584     \\
$x_{2B}  $                  &0.785                            & 0.799     \\
$x_{2V}  $                  &0.688                            & 0.705     \\
$x_{2R}  $                  &0.592                            & 0.611     \\
$y_{1B}  $                  &0.294                            & 0.294     \\
$y_{1V}  $                  &0.294                            & 0.294     \\
$y_{1R}  $                  &0.294                            & 0.294     \\
$y_{2B}  $                  &0.283                            & 0.252     \\
$y_{2V}  $                  &0.290                            & 0.282     \\
$y_{2R}  $                  &0.291                            & 0.285     \\
$T_{1}   $ (K)              &7178                             & 7178      \\
q ($M_2/M_1$ )              &0.984                            & 0.984     \\
$\Omega_{in}$               &3.7239                           & 3.7239    \\
$\Omega_{out}$              &3.1880                           & 3.1880    \\
$i$   (deg)                 &62.80 $\pm1.01$                  & 57.73$\pm0.48$     \\
$T_{2}$ (K)                 &7003  $\pm36$                    & 6631  $\pm49$      \\
$\Omega_{1}$                &3.6419$\pm0.0142$                & 4.6120$\pm0.0472$  \\
$\Omega_{2}$                &3.6419$\pm0.0142$                & ---                \\
$L_{3}/(L_{1}+L_{2}+L_{3})$ ($B$) &$0.102\pm0.004$            & $0.026\pm0.002$ \\
$L_{3}/(L_{1}+L_{2}+L_{3}$) ($V$) &$0.088\pm0.004$            & $0.013\pm0.002$ \\
$L_{3}/(L_{1}+L_{2}+L_{3}$) ($R$) &$0.074\pm0.004$            & $0.002\pm0.002$ \\
$r_{1}$   (pole)            &0.3679$\pm0.0019$                & 0.2730$\pm0.0035$  \\
$r_{1}$   (side)            &0.3883$\pm0.0023$                & 0.2788$\pm0.0038$  \\
$r_{1}$   (back)            &0.4247$\pm0.0034$                & 0.2871$\pm0.0042$  \\
$r_{2}$   (pole)            &0.3652$\pm0.0019$                & 0.3548             \\
$r_{2}$   (side)            &0.3854$\pm0.0023$                & 0.3726             \\
$r_{2}$   (back)            &0.4219$\pm0.0034$                & 0.4035             \\
degree of overcontact (f)    &17.4\%$\pm2.5\%$                     & ---                   \\
Residual                    &0.0091                           & 0.0111     \\
\hline\noalign{\smallskip}
 \end{tabular}
 \end{center}
 \end{table}

\begin{figure}[h!!!]
\centering
\includegraphics[width=9.0cm,angle=0]{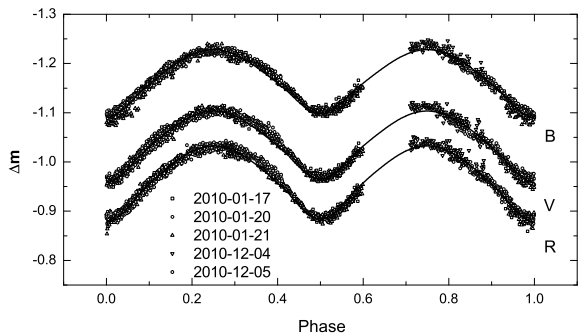}
\begin{minipage}[]{85mm}
\caption{Same as the top panel of Figure 1. But the solid curves
represent the theoretical light curves computed with the
parameters in Table 2.}
\end{minipage} \label{Fig2}
\end{figure}

\section{Discussion and Conclusions}
\label{sect:discussion}

The investigation on the new multicolour light curves indicated
that FT UMa is an evolved contact binary. The system shows two
atypical properties: the mass ratio close to unity and the small
photometric amplitude.

The typical mass ratios of contact binaries are between 0.2 and
0.5 (Gettel et al. 2006). The contact binaries with large mass
ratios, especially those with unit mass ratios, can help us
understand the evolution of contact binaries, and study the link
between A- and W-subtype W UMa binaries (Li et al. 2008). In the
case of FT UMa, two evolved components have almost the same masses
and radii, and therefore little mass exchange. If both of
components can evolve onto the subgiant stage, the system will
coalesce directly into a single star .

In addition to FT UMa, the mass ratios as high as unity can be
seen in other five contact binaries: V701 Sco (the spectroscopic
mass ratio $q_{sp}=0.99$: Bell \& Malcolm 1987) and V753 Mon
($q_{sp}=0.970$: Rucinski et al. 2000), CT Tau (the photometric
mass ratio $q_{ph}=1.00$: Plewa \& W{\l}odarczyk 1993), V803 Aql
($q_{ph}=1.00$: Samec et al. 1993), and WZ And ($q_{ph}=1.00$:
Zhang \& Zhang 2006).

Moreover, the large total mass of $2.95~M_{\odot}$ and the large
mass ratio of $q=0.984$ are consistent with the fact that the mass
ratio of the W UMa-type systems increases with the increase of
their total mass (Li et al. 2008).

Generally, a close third component, orbital inclination and the
relative geometrical size of two components can affect the
amplitude of photometric variations. FT UMa has an orbital
inclination of $62.8^{\circ}$ and almost identical components in
size, and therefore show a relatively low photometric amplitude,
$\sim 0.15$ mag. Pribulla et al. (2009) concluded that a third
component contributes about an half light of the system and
reduces the amplitude of photometric variations. But our solution
suggested that the light contribution is about $10\%$. Just as
noted by Pribulla et al. (2009), the center-of-mass velocity of
the close pair was changeless during their observational run,
compared with the variable velocity of third component. This
indicates that the mass of the third component is much smaller
than the total mass of the binary pair, suggesting that the light
dilution of a third component is negligible. In order to confirm
the light and mass of the third component, investigation on
long-term orbital period and radial velocity are needed.

\normalem
\begin{acknowledgements}
We thank an anonymous referee for some useful suggestions. This
work is supported by Natural Science Foundation of Shanxi Normal
University (No. ZR09002).
\end{acknowledgements}

\label{lastpage}

\end{document}